# Bucky-Corn: Van der Waals Composite of Carbon Nanotube Coated by Fullerenes

*Leonid A. Chernozatonskii,[a*] Anastasiya A. Artyukh,[a] Victor A. Demin,[a] Eugene A. Katz[b,c*]*

a. Emanuel Institute of Biochemical Physics, RA S, Moscow,119334 Russia

b. Dept. of Solar Energy and Environmental Physics, The Jacob Blaustein Institutes for Desert Research (BIDR), Ben-Gurion University of the Negev, SedeBoker Campus 84990, Israel

c. Ilse Katz Institute for Nanoscale Science & Technology, Ben Gurion University of the Negev, BeerSheva 84105, Israel

Corresponding author, e-mail address: keugene@bgu.ac.il, chernol-43@mail.ru

ABSTRACT: Can C60 layer cover a surface of single-wall carbon nanotube (SWCNT) forming an exohedral pure-carbon hybrid with only VdW interactions? The paper addresses this question and demonstrates that the fullerene shell layer in such a bucky-corn structure can be stable. Theoretical study of structure, stability and electronic properties of the following bucky-corn hybrids is reported: $C_{60}$ and $C_{70}$ molecules on an individual SWCNT, $C_{60}$ dimers on an individual SWCNT as well $C_{60}$ molecules on SWNT bundles. The geometry and total energies of the bucky-corns were calculated by the molecular dynamics method while the density functional theory method was used to simulate the electronic band structures.

**TOC GRAPHICS**

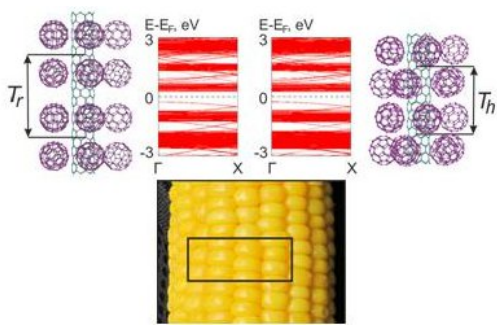

# 1. Introduction

Organic photovoltaics (OPV) has been suggested as an alternative to conventional photovoltaics due to their light weight, mechanical flexibility, and processability of large-area, low-cost devices. In particular, intense research is directed towards the development of OPV with a bulk heterojunction (BHJ) where donor-type conjugated polymers (hole conducting) and acceptor-type (electron conducting) fullerenes or fullerene derivatives.[1-3]

Upon illumination, light is absorbed by the conjugated polymer resulting in the formation of a neutral and stable excited state on the polymer chain. Free carriers can be generated by exciton dissociation at a donor-acceptor interface, leaving the electron on the acceptor (fullerene in this case) and the hole on the conjugated polymer donor. Efficient charge generation requires, therefore, that the donor and acceptor materials form interpenetrating and continuous networks, "phase separated" on the scale of the exciton diffusion length: ≤10 nm.[4] Following the exciton dissociation into free carriers, the electrons and holes are conducted through the respective semiconductor moieties (fullerene percolation network for electrons, and conjugated polymer chains for holes) towards the respective electrodes.

One of the factors limiting the OPV efficiency is the poor electronic hopping-transport in the fullerene channel of the BHJ cells, resulting in a very high fullerene percolation threshold of about ~50%. The high fullerene content improves charge transport at the expense of sunlight absorption (as fullerenes exhibit much lower absorption in the visible range of the spectrum as compared to conjugated polymers).[5]

Recently carbon nanotubes (CNT)[6] were suggested as a favourable alternative for fullerenes in BHJ solar cells.[7-10]

The main advantages of CNT, as compared to fullerenes, include their wide range of diameter-dependent band gaps (from 0 to 1.2 eV) that can match the solar spectrum, and reduced carrier scattering for hot carrier transport resulting in a near-ballistic transport along the tube.[6]. Additional major advantages are related to the nanotube geometry: the high aspect ratio (length-to-diameter) >$10^3$ of SWNT yields percolation thresholds below 0.1 weight percent (wt%) offering high-mobility pathways for electron transport; and their high surface area (~1600m2/g) provides high CNT/polymer interfacial area for charge separation.[11] Therefore, both critical processes governing the operation of conjugated polymer BHJ solar cells, i.e. charge separation and charge transport, were expected to be enhanced in CNT/conjugated polymer PV.

In spite of the great initial expectations, currently reported efficiencies in CNT-based BHJ solar cells are well below 1%.[7-10] A comprehensive review of the factors limiting efficiency of such devices can be found in ref. 12. The main of these factors is inefficient electron transfer at the CNT (acceptor)/polymer (donor) interface.[13]

The difficulties in utilization of CNT as electron acceptors with yet the need to replace fullerenes by more favourable network for electron transport have led to the suggestion of using CNT exclusively for charge transport in a 3-component system: CNT-fullerene-conjugated polymer.[14-20] It was suggested that by combining both low concentrations of fullerenes and CNT one may be able to still utilize fullerenes for efficient charge separation while CNT with their superior electron transport properties and their low percolation threshold would serve for setting high-mobility pathways for electron transport. While of great potential, the morphology of the combined three-component system is very hard to control due to the very strong interactions within both fullerene and CNT components.

From this point of view, the ideal CNT-core/fullerene-shell hybrid structure (if exists) may constitute an ideal photoactive component for BHJ OPV.

Hybrid structures consisting of fullerenes and CNT are under intense research attention during last years. Reviewing these studies Vizueteet al.[21] suggested to classify such structures as endohedral and exohedral hybrids. In the endohedral hybrid $C_{60}$ fullerenes, for example can be enclosed into CNT to form the so-called carbon nanopeapod ($C_{60}$@CNT), where the interaction between fullerenes and CNTs is of van der Waals type.[22] Unfortunately, the reported exohedral hybrids (that are required for the efficient OPV) include only either carbon nanobuds, i.e. covalently bonding C60 to the CNT sidewall[23,24] or interconnection of fullerenes and CNT through their fictionalization.[25-28] However, covalent sidewall chemical functionalization is known to disrupt the conjugated π-systems of CNT and in turn deterioration of their unique electronic properties.[6]

$C_{60}$ fullerenes on a flat metal crystal surface are known to form a Van der Waals (VdW) close-packed layer[29] mostly with hexagonal symmetry (Fig. 1a). Recently this structure was also found for $C_{60}$ layer on a graphite surface.[30] Here, we address a question if $C_{60}$ layer can cover a surface of single-wall CNT forming an exohedral pure-carbon hybrid in which SWCNT core is covered by fullerene shell with only VdW interactions. We demonstrate that the fullerene shell layer in such a bucky-corn structure can be stable and form a quasi-1D crystal similar to a fullerene nanowire.[31] We report theoretical study of structure, stability and electronic properties of the following bucky-corn hybrids: $C_{60}$ and $C_{70}$ moleculeson an individual SWCNT, $C_{60}$ dimers on an individual SWCNT as well $C_{60}$ molecules on SWNT bundles. The hybrids with multi-wall CNT can be simulated in a similar way however it requires much larger computational resources.

The geometry and total energies of the bucky-corns were calculated by the molecular dynamics (MD) method using the GULP program.[32] In these calculations C-C bonds were parameterized on the basis of Brenner potentials while intermolecular interactions were simulated using Lennard-Jones potentials. The density functional theory (DFT) method[33] implemented in the SIESTA software packet[34] was used to simulate the electronic band structures.

## 2. Results and discussion
### 2.1. $C_{60}$ shell on individual SWNT
#### 2.1.1 Structure and stability

We hypothesized that distances $d_{VdW}$ between SWCNT and $C_{60}$ molecules as well as between adjacent $C_{60}$ molecules are approximately equal to those in high temperature (T > 260 K) VdW fullerite phase. In the latter, $C_{60}$ molecules are ordered in an *fcc* crystal structure with lattice constant of 1.417 nm and a nearest-neighbor $C_{60}$-$C_{60}$ distances of 1.002 nm, and rotated rapidly (period of ~ 10 psec) with three degrees of rotational freedom.[35] For all simulated SWCNT and $C_{60}$ hybrids our MD calculations demonstrated similar phenomenon: at room temperature T=300K $C_{60}$ molecules are rotated with three degrees of freedom. For example, in (5,5)CNT@$C_{60}$-h(6,0) bucky-corn the period of such rotation was calculated to be ~30 psec at T=300K. Accordingly, one can consider $C_{60}$ molecules as equivalent "balls" and apply to their close-packed shell a simple geometrical approach of the net of points (centers of molecules) similar to the flat net of carbon atoms unfolded from the cylindrical SWCNT surface (Fig. 1).

One can simply estimate a diameter of such a shell cylinder as

$$D \approx D_{SWCNT} + d_f + 2d_{VdW}, \quad (1)$$

where $D_{SWCNT}$ and $d_f$ are diameters of SWCNT and the $C_{60}$ molecule, respectively; while $d_{VdW}$ is a distance between them ($d_f \approx 0.7$ nm, $d_{VdW} \approx 0.3$ nm).

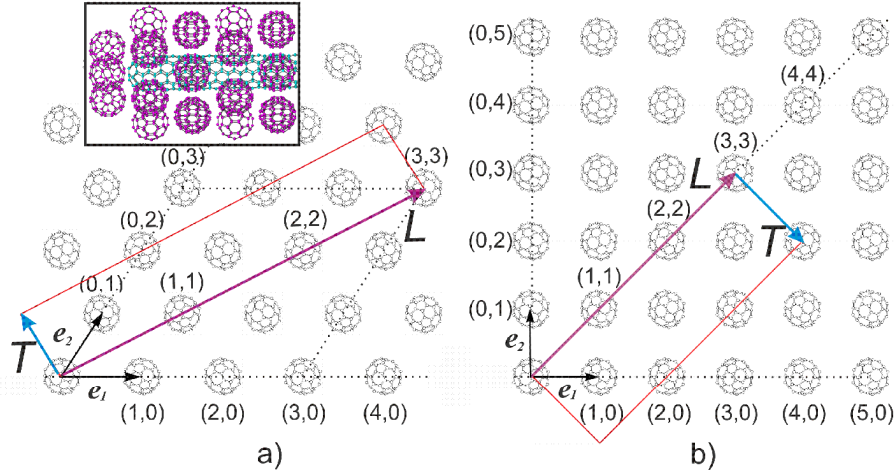

**Figure 1.** Close-packing of $C_{60}$ molecules on the plain with the chiral vector $L(k, l)$: a) the densest hexagonal structure (inset shows an example of (12,0)CNT@$C_{60}$-h(8,0)), b) rhombic structure

It should be noted that at low temperatures (T < 240 K) various combinations of mutual arrangement of carbon atoms on the adjacent SWCNT cylinder and $C_{60}$ molecules are possible (similar to the corresponding phenomenon in the fullerite crystal[35]).

Each fullerene cylindrical layer $C_{60}(k,l)$ can be unambiguously described by the chiral vector $L(k,l)$ of its flat net which is equal to the perimeter of the circular cross section of this cylinder, using a grid of dots (fullerenes centers), similar to that of carbon atoms for SWCNT.[6] Figure 1 shows the simplest examples of such grids, namely hexagonal (h) and rhombic (r) packing of centers of fullerenes, spaced at a distance $d_f + d_{VdW} \approx 1$ nm (as well as in the fullerite[35]) with the unit cells defined by two unit vectors.

Inset in Figure 2a depicts a possible mixed structure consisted of strips of the $C_{60}$ pairs with a rectangular unit cell with four fullerenes. It is interesting to note that such a structure can often be found on the natural corn on the cob (Fig. 2b).

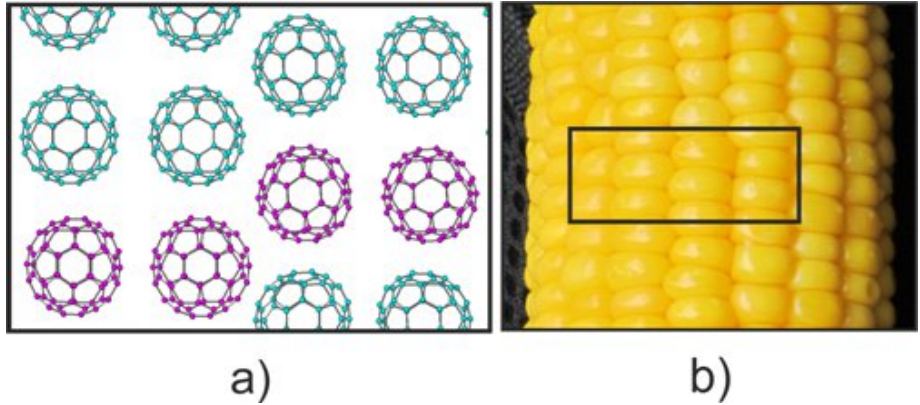

**Figure 2.** A mixed packing with twin (paired) $C_{60}$ rows shifted relative to each. Model (a) of selected area of the natural corn on the cob (b).

Using the geometry of the "hexagonal" layer on the plane (Fig.1a) one can define the relationship between the diameter of this layer $D_h(k,l)$ and the distance between fullerenes $d_{ff}$. The diameter and the period of $T_h$ superlattice of the $C_{60}$-h(k,l) layer can be estimated as:

$$D_h(k,l) = \frac{L(k,l)}{\pi} = \frac{d_{ff}}{\pi}\sqrt{k^2 + kl + l^2}; \quad T_h = d_{ff}\frac{\sqrt{3}\sqrt{k^2+kl+l^2}}{?} \quad (2)$$

where $k$ and $l$ ($l < k$) are integers determined by coordinates of the grid (Fig.1a).

Similarly, it is easy to estimate the corresponding parameters for "rhombic" fullerene layer $C_{60}$-r(k, l) using the grid coordinates in Fig.1b ($D_r$, $T_r$).

For calculation of cohesive energy for formation of various bucky-corns, the periodic boundary conditions were chosen. The cohesive energy was calculated as:

$$E_b = \frac{E_{tot} - (E_{CNT} + E_f N_f)}{N_{atoms}} \quad (3)$$

where $E_{tot}$ is the total energy of the system, $E_{CNT}$ is energy of the corresponding region of isolated nanotube, $N_f$ is a number of fullerene molecules in the shell, $E_f$ is energy of an isolated fullerene molecule, $N_{atoms}$ is a number of atoms in the calculated cell.

Some results of such calculation are summarized in Figure3. Minimum energy was calculated to the structure with carbon nanotube (5.5)with a diameter of 0.69 nm that is close to the diameter of the fullerene (~ 0.7 nm). This is due to the fact that in such structure (5,5)CNT@$C_{60}$-h(6,0) the fullerene-fullerene and fullerene-nanotube distances have values closest to the intermolecular distances in the $C_{60}$ crystal. For the same nanotubes hexagonal fullerene coating is more favorable since this is the most dense type of packaging.

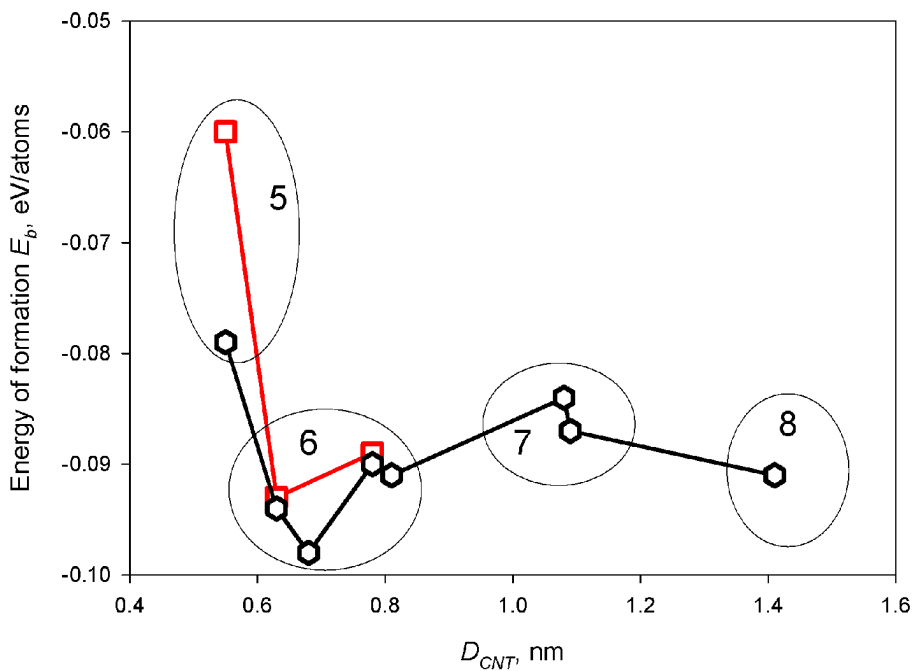

**Figure 3.** Cohesive energy for formation of various bucky-corns. Red square and black hexagons mark values for the rombic and hexagonal bucky-cornes, respectively. Number in the ellipse indicates the number of fullerenes in the ring.

The values of the bucky-corn cohesive energy are shown to be very close to that for the fullerite fcc crystal (-0.113 eV/atom). This fact indicates that bucky-corns can be stable under the conditions of fullerite stability. Additional layers of fullerenes (for example, two-shell structure (5,5)CNT@$C_{60}$-h(6,0)@$C_{60}$-h(11,0)) make the "corn" even more stable.
.

### 2.1.2. Electronic band structures

Here we demonstrate calculation results of electronic band structures of bucky-corns consisting of achiral fullerene shell on achiral CNT with a diameter of ~1 nm (nanotubes of this diameter are the most abundant among the synthesized SWCNTs[36]). Our choice is dictated by a smaller number of atoms in the super-cell of such structures as well as by the simplest selection of commensurability of periods of fullerene shell and core nanotube. The number of atoms in the super-cell significantly increases for chiral systems and their simulation requires a large array of computing.

Eq. (1) indicates, for instance, that the densest packing of the six fullerenes with centers lying in a circle at the plane perpendicular to the CNT axis is realized for the SWCNT with diameter of 1 nm ((8,0), (9,0) (5,4), (5,5)). In the case of the zigzag CNT, 4 cells are packed within one period of the fullerene "hexagonal" shell layer $T_h(6,0) \approx 1.7$ nm (doubled period $2T_r(6,0) \approx 2.1$ nm with 5 cells for "rhombic" shell). Figures 4a and 4d depict the structures of "rhombic" and "hexagonal" hybrids $(8,0)CNT@C_{60}$-x(6,0), where x = r, h. Cell of each of two these structures $((8,0)CNT@C_{60}$-r(6,0) or $(8,0)CNT@C_{60}$-h(6,0)) comprises a part of the nanotube (8,0) of 80 or 64 carbon atoms and 12 fullerene molecules (800 or 784 carbon atoms in the super cell of bucky-corn), respectively. After the optimization, the translation parameter $T$ along the axis of r and h structures was equal to 2.13 and 1.70 nm, respectively.

The electronic band structure is found to be almost independent of the location of fullerenes in the shell. It contains minizone associated with the removal of degeneracy of the levels of $C_{60}$ truncated icosahedron due to its interaction with carbon nanotubes and other adjacent fullerenes (as in the fullerite[17]). The band gap $E_g$=0.52 eV for this bucky-corn structure with a semiconductor CNT ($E_g$ = 0.58 eV) is slightly diminished (Fig. 4b and c). Minizones are sufficiently wide (the widths are ~0.4 eV and ~0.8 eV for the first minizones above and below the Fermi level ($E_F$), respectively. First minizones are separated by energy of ~0.9 eV. The width of minizones and energy gap are determined by fullerene shell and type of nanotube, respectively.

The calculation for(5,5)CNT (with a period $T_{CNT(5,5)}$=0.246 nm) surrounded bya shell $C_{60}$-h(6,0) (bucky-corn (5,5)CNT@$C_{60}$-h(6,0) with a period $T_r$=0.99 nm ≈ $4T_{CNT(5,5)}$ = 0.98 nm) reveled an electronic structure with arange of the same wideminibands on which almostunchangedelectronic band structures of the metalnanotube is superimposed. To givean idea ofthe transformationof the degenerated levelsof $C_{60}$fullerene into theminizonesdue tothe contribution ofa large number ofbranchesofthe deformedandinteractingfullerenes.

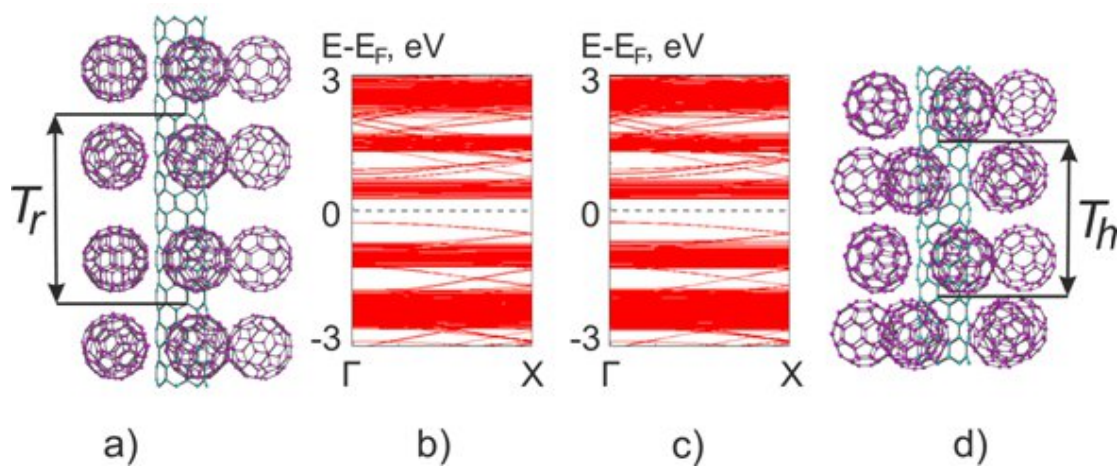

**Figure 4.** Structures (a, d) and electronic band structures (b, c) of bucky-corns (8,0)CNT@$C_{60}$-r(6,0) and (8,0)CNT@$C_{60}$-h(6,0), respectively.

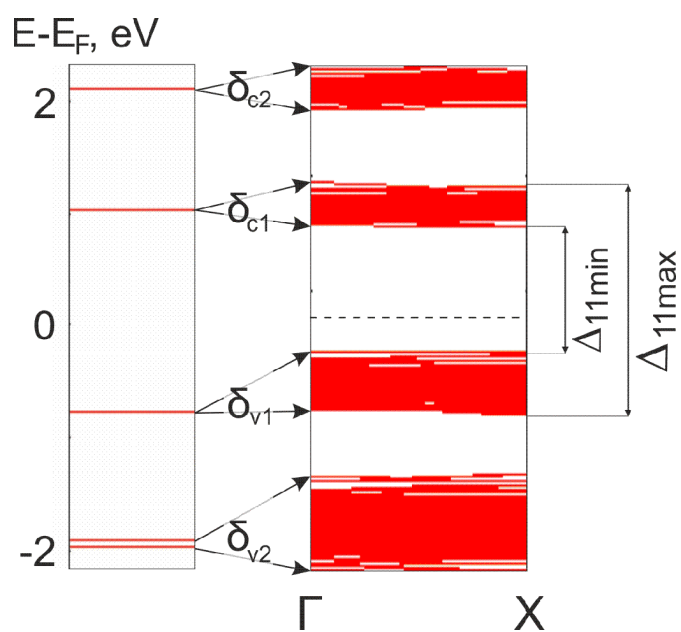

**Figure 5.** Electronic band structures of fullerene $C_{60}$ and the "hexagonal" fullerene shell layer $C_{60}$-h(5,0) in the bucky-corn (7,0)CNT@$C_{60}$-h(5,0).

Figure 5 compares the electronic band structures of fullerene $C_{60}$ and the "hexagonal" fullerene shell layer $C_{60}$-h(6,0) in the bucky-corn (5,5)CNT@$C_{60}$-h(6,0). For this structure, the minizones near to the Fermi level have the widths of $\delta_{v1}=0.51$ eV, $\delta_{v2}=0.78$ eV, $\delta_{c1}=0.39$ eV, $\delta_{c2}=0.42$ eV. Thus, the photoexcitations of electrons from the minizones below the $E_F$ up to those above the Fermi level could occur with the energy of $\Delta_{11min}=1.08$ eV, $\Delta_{11max}=1.97$ eV, $\Delta_{22min}=3.15$ eV, $\Delta_{22max}=4.35$ eV. The transition between minizones $\delta_{0i}$ and $\delta_{i0}$ determines by interval [$\Delta_{ii\,min}$; $\Delta_{ii\,max}$].

The range of energies from $\Delta_{11min}$ to $\Delta_{11max}$ corresponds to the absorption of light with wavelength of 630 - 1150 nm (the red and infrared light), while energy range, [$\Delta_{22min}$; $\Delta_{22max}$] controls the absorption of the violet and ultraviolet light (285-390 nm). The number of branches of the electronic band structure between the mini-bands is also increased, as compared with the electronic band structure of the pristine SWCNT. This effect is observed due to the fact that the increased period of super-lattice (in comparison with that for the pristine SWCNT) results in reduction of the Brillouin zone. Light absorption by bucky-corns with various types of nanotubes will be different. However, the ranges of allowed transitions between the fullerene mini-bands vary just slightly with increase in the diameter of carbon nanotubes and the number of fullerenes in the super-lattice. This is evident from the comparison of the widths and positions of mini-zones in the electronic band structures of the bucky-corns $(8,0)CNT@C_{60}$-h$(6,0)$ (Fig.4), $(14,0)CNT@C_{60}$-h$(7,0)$ and $(8,8)CNT@C_{60}$-h$(7,0)$ (Fig.6).

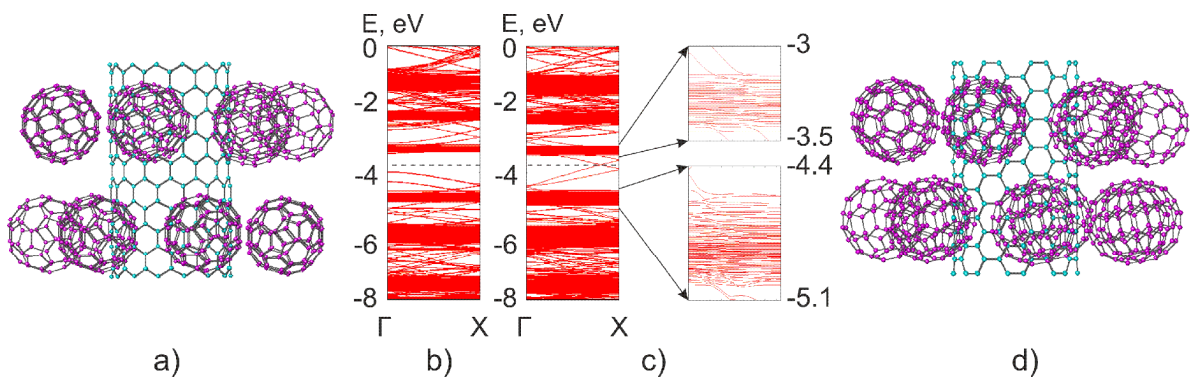

**Figure 6.** General views and electronic band structures of hexagonal bucky-corns $(14,0)CNT@C_{60}$-h$(7,0)$ (a and b) with semiconductor single-walled nanotube and $(8,8)CNT@C_{60}$-h$(7,0)$ (c and d) with metallic single-walled nanotube.

## 2.2. Bucky-corns with bundles of SWNTs

Figure 7 demonstrates the structure and electronic band structure of three fullerene rows (chains) on the bundle of three (5,5)CNTs. The super-lattice consists of 420 atoms. The period along the bundle axis is equal to 0.99 nm.

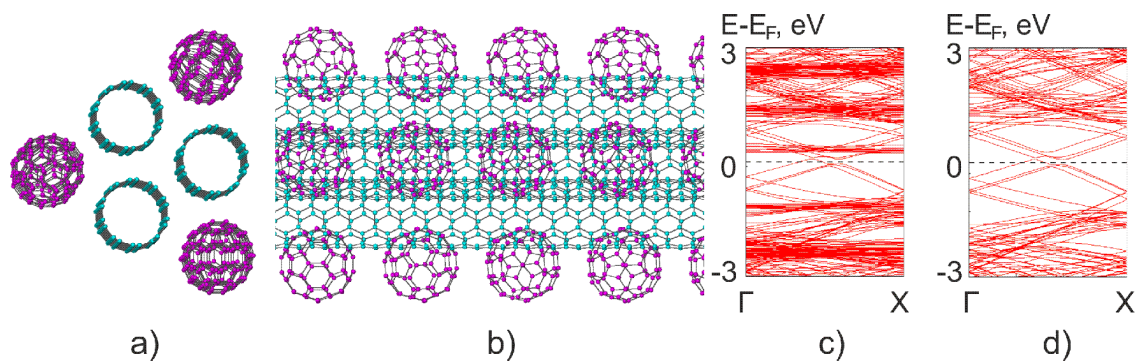

**Figure 7.** Top and side views (a, b) and electronic band structure (c) of three fullerene rows on the bundle of three (5,5)CNTs and electronic band structure (d) of the three (5,5)CNTs in bundle for comparison.

As one can see in Figure 7c the branches of the electronic band structure related to the nanotube wave functions are tripled because of the molecular bonds among the (5,5) nanotubes. This creates an additional range of possible electronic transitions in the interim between the fullerenes' miniband parts of the band structure. The structure is also of metallic type, as well as its nanotube components taken separately (see and compare electronic band structures in fig. 7 c and d).

Covering the CNT bundle by a dense fullerene shell (Fig 8a and b) increases the cohesive energy and the minibands' population. It is natural to expect that when a large number of fullerenes presents in a solution, they will be attracted to the "expedient" places between the fullerene rows and nanotubes. In the case of three CNTs ($D_{CNT}$~0.7 nm) bundle it is possible to form shell of 9 fullerene rows around the bundle (Fig. 8a). This shell layer can be "rhombic" structure if it is deployed on a plane (as in Fig.1b). This structure is preferable energetically than the "unfinished" shell shown in Fig. 7 (by 0.003 eV/atom). Of course, in the case of the bundle of nanotubes with diameters different from that of fullerene, the shell may have a structure other than "rhombic".

We have simulated a corn structure (Fig. 8a) with a bundle of seven (5,5)CNTs covered a shell layer of twelve fullerene chains 7(5,5)CNT@$C_{60}$-r(12,0). Its cohesive energy was found to be just a bit lower (by 0.0001eV/atom) than that for the bucky-corn 3(5,5)CNT@$C_{60}$-r(12,0) (Fig. 8b).

Thus, one can conclude that bucky-corns with the increased number of nanotubes in a bundle and fullerene molecules in a shell are still as energetically stable as the fullerites.

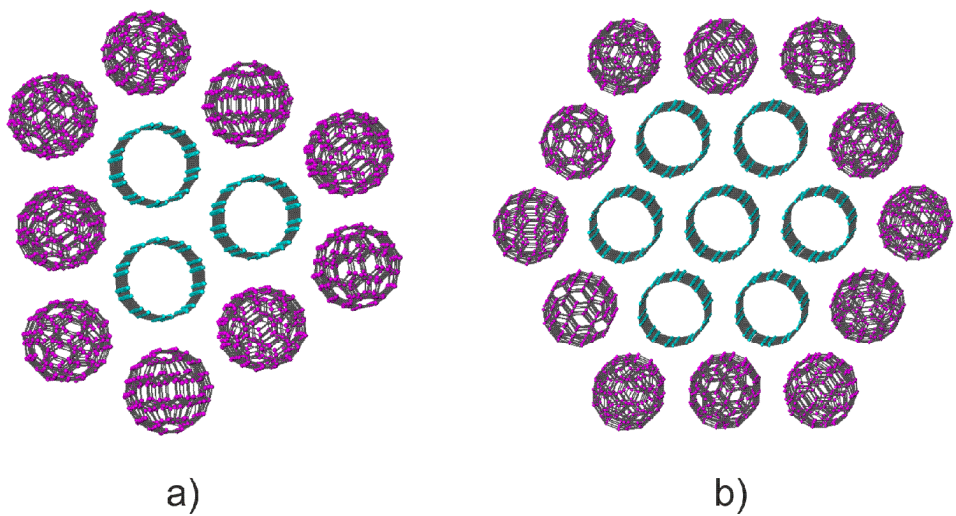

a)          b)

**Figure 8.** Bucky-corns on the three CNTs (3(5,5)CNT@$C_{60}$-r(12,0)) (a) and seven CNTs (7(5,5)CNT@$C_{60}$-r(12,0)) (b).

### 2.3. Bucky-corns with fullerene dimers and polimer chaines

Fullerenes are known to be photopolymerized under pressure, ion and electron irradiation or UV illumination.[37] $C_{60}$ dimers produced in the first steps of this process have an enriched energy spectrum (in comparison with the pristine $C_{60}$).[35] Here, we address a question if it can help to broaden the "fullerene" minizones in the bucky-corn electronic band structure and accordingly to achieve the more effective light absorption. Since the main mechanism of $C_{60}$ photopolymerization is a [2+2] cycloaddition of two $C_{60}$ molecules,[35] we will consider bucky-corns with a shell of the [2+2] dimers on the SWCNT with a diameter of 1 nm. On the cylinder of such a diameter (nanotubes (8,0) or (5,5)) a dense packing consists of six fullerene molecules with centers lying in a circle in the plane perpendicular to the tube axis (for example, non-chiral shell d6$C_{60}(\alpha)$).

Therefore, the MD simulation was performed for shell structures with three possible locations of the [2+2] dimers around such a tube (with annealing at 300 K).

Dimerisating results in fullerene resizing. The molecules are elongated along the covalent and compressed in the perpendicular direction.[35] Therefore the calculations carried out with the same translation period can just show a "qualitative" trend. In bucky-corn modeling a disparity between periods of nanotubes and the shell of fullerene dimers arises. Translation period for the shell of the dimers with angle α = 0° (Fig. 9 a) is described as:

$Z_d = L_d + d_{VdW}$,  (4)

where $L_d = 2D_{def\,f} + l_{C-C}$ is a dimer length equal to the sum of twice the diameter of the deformed fullerene and length of covalent C-C bond, $d_{VdW}$ is intermolecular distance. To estimate the corresponding value for the fullerene shell one should get twice the translation period $2T_r(6,0) = 2(D_f + d_{vdW})$. It is obvious that $2T_r > Z_d$, since the difference between the diameters of the fullerene and the deformed fullerene in the dimer does not compensate for the difference of the covalent bond length and the intermolecular distance. However, we have a restriction on the choice of the computational cell (the translation parameter of the nanotube fragment). We cannot change the fullerene diameter or the dimer length and the size of the computational cell must be reasonable. Therefore to model the bucky-corn we assume a small change in the intermolecular distance. In case of fullerene bucky-corn the disparity between periods of the nanotube and the fullerene shell does not lead to significant changes in the intermolecular distance. However, in the structure of the dimer bucky-corn (such as $(8,0)CNT@d6C_{60}(α = 0°)$) the distance between the dimers in the neighboring calculated cells is high (0.41 nm). Accordingly, the cohesive energy is much higher than that for the structures with a different arrangement of dimers (α = 90°, ≈60°) – Fig. 9b and c.

We calculated the cohesive energy per atom in various dimer configurations using eq. (3). For the CNT (8,0) the shell configuration with three dimers located along the circle at an angle α=90° to the tube axis (Fig. 9c) is the most favorable ($E_b$ = - 0.18 eV/atom). The shell of three dimers arranged around the ring at an angle α≈60° (Fig 9b) constitutes the next stable configuration ($E_b$ = - 0.17 eV/atom). A shell of six dimers elongated along the nanotube axis (Fig 9c) was revealed to be less favorable ($E_b$ = -0.02eV/atom), since the distance between the neighboring dimers along the tube is high for this configuration. This configuration isn't stable in room temperature T=300K. Angle α is changed to ≈60°.

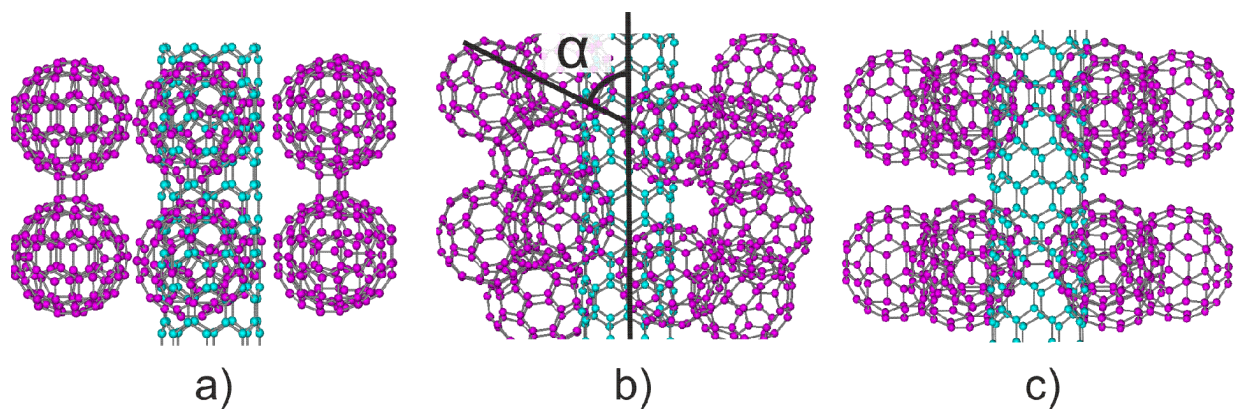

**Figue. 9.** Bucky-corns with the (8,0)CNT and the shell of the dimers located along the circle at an angle α to the tube axis: (a) α=0°, (b) α≈60°, (c) α=900°.

In order to compare directly electronic band structures of the bucky-corns with shells consisted of the fullerene dimers and pristine fullerene molecules, we have simulated another type of the corn, namely (5,5)CNT@d6C$_{60}$(α) (Fig. 10). A similar result was obtained for the structure of CNT(8,0)@d6C$_{60}$ (α≈45°). The fact that mini-zones in the "dimer" shell are much wider and more populated is in evidence. The corn bandgap in this case is determined by the nanotube while mini-zones are wider than those in a similar structure without dimers ((5,5)CNT@C$_{60}$-h(6,0)).

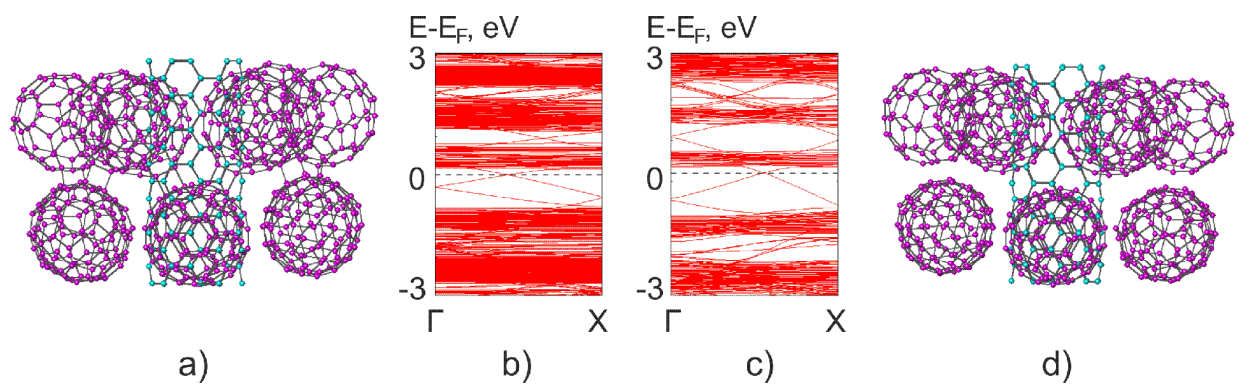

**Figure 10.** The "dimer" corn (5,5)CNT@d6C$_{60}$(α≈45°) (a) and its electronic band structure (b) in comparison with the electronic band structure (c) of the corn CNT(5,5)@C$_{60}$-h(6,0) with pristine fullerenes (d).

With longer (or/and stronger) polymerization process more fullerene molecules are covalently bonded in the polymerized chains. Because of the period disparity of the nanotubes and the fullerene chains we modeled stable bucky-corns of finite sizes.

Let us show one configuration of stacking of linear fullerene chains along the axis of the nanotube, in which fullerene molecules are connected by [2+2] cycloaddition (Fig. 11a). This [2+2] configuration is often experimentally.[35] The binding energy calculation for such a structure ((5,5)CNT@6chC$_{60}$) showed that it is more favorable energetically than the bucky-corn with the dimers (CNT(5,5)@d6C$_{60}$($\alpha \approx 45°$)). C$_{60}$ molecules are stretched along the chain losing a spherical shape. Their size along the tube increases and the perpendicular size decreases. This leads to an increase in the distance between the two chains in the corn structure CNT(5,5)@ 6 ch C$_{60}$.

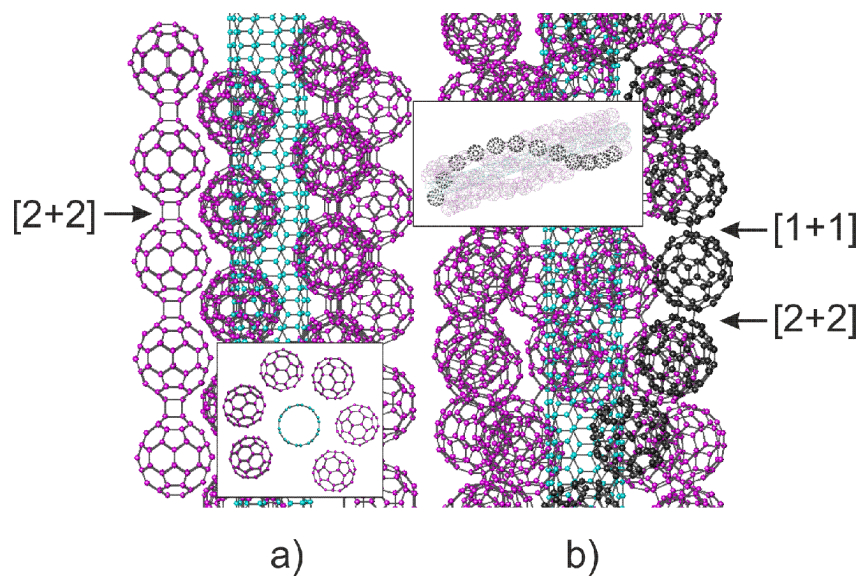

**Figure 11.** Straight chains of covalent-bonded [2+2] fullerenes on the (5,5)CNT (a) and (inset - its side view). Chain of dimers [2+2], linked to each other by one covalent bond forming a continuous helical chain polymers around (5,5)CNT (b) and (inset - its general view).

Other configurations of polymerized C$_{60}$ with chiral chains are also possible. For example, Figure 11b shows a chain of three dimers [2 + 2], linked to each other by one exopolyhedralbond.[39] Structural optimization for the corn with six spirals (similar to that shown by Figure 11a) has demonstrated a higher value of cohesive energy than that for the corresponding corn with unbounded dimers ((5,5)CNT@d6C$_{60}$ ($\alpha \approx 45°$)) by ~0.02 eV/atom. This indicates the greater resistance of the former one to high temperatures. It is logical to assume that the covalent interaction of fullerene spirals will not lead to a substantial change in the electronic properties compared to its predecessor ((5,5)CNT@d6$C_{60}$ ($\alpha\approx45^{o}$)).

### 2.4. $C_{70}$ bucky-corns (SWNT in $C_{70}$-shell)

Fullerene $C_{70}$ is known to have less symmetry that $C_{60}$ molecule. This is the reason why the levels of the electron spectrum of the former are less degenerated than that of the latter. It was therefore interesting to consider the bucky-corn structure, replacing the $C_{60}$ by $C_{70}$. The corn configurations in which the axes of $C_{70}$ molecules are parallel to the nanotube axis are found to be optimal for all CNT in $C_{70}$ shell structures.

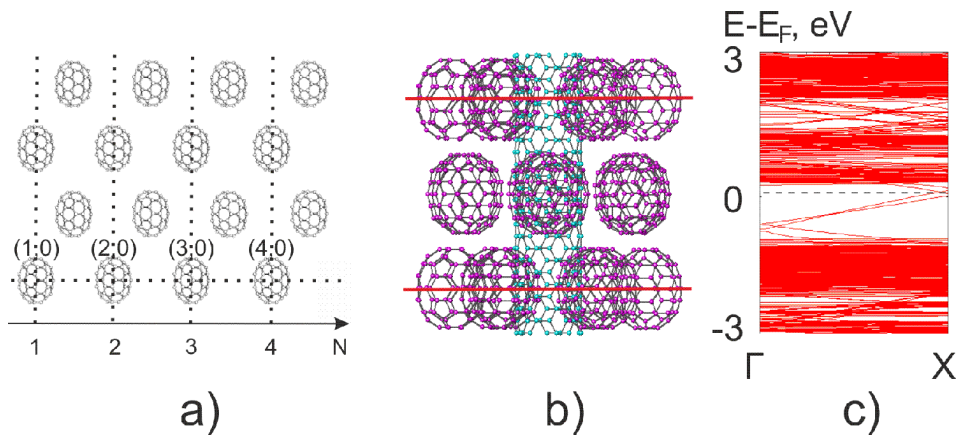

**Fig. 12.** Bucky-corn (5,5)CNT@$C_{70}$-h(6,0): plane net for non-chiral $C_{70}$ shell (a); side view of bucky-corn (5,5)CNT@6$C_{70}$-h(6,0)(b) (the unit cell is shown by a red lines), its electronic band structure (c).

Simplest non-chiral configuration is that in which the ring of integer number of $C_{70}$ molecules is in the plane, perpendicular to the nanotube axis. The plane net of this shell cover is centered rectangular lattice (Fig. 12a). The diameter of the fullerene ring is calculated by eq. (1) using the corresponding $C_{70}$ parameters. Figure 12b shows a model of the CNT (5,5) covered by a hexagonal shell of six $C_{70}$ molecules in the ring. The calculated cell for this bucky-corn consists of 1000 carbon atoms. Translation parameter along the nanotube axis is equal to 1.96 nm. Contrary to the the electronic band structure of CNT in $C_{60}$-shell that for CNT in $C_{70}$-shell layer in the electronic structure does not consist of the distinct minibands (Fig. 12c). This should result in the better electron-light interaction in a wide wavelength range. Conductivity type of the corn is defined by the core nanotube (semimetallic in the case of (5,5)CNT).

In other cases, the axis of $C_{70}$ molecules will form an angle to the nanotube axis, which complicates the modeling of such CNT in $C_{70}$-shell structures.

## 3. Conclusions

We propose a new carbon van der Waals composite nanostructure consisting of a nanotube covered by close-packed shell of fullerene molecules. The geometry and total energies (stability) of the of the fullerene shell layer in such *bucky-corn* structures were calculated by the MD method while the DFT method was used to simulate its electronic band structure. Value of energy gap is controlled by the type of core carbon nanotube. Width of mini-zones is defined by fullerene shellsince fullerenes are deformed in bucky-corns and they lost symmetry and degeneracy of electronic levels. In addition to CNTin $C_{60}$- or $C_{70}$-shell, bucky-corns with CNT bundles as well the shells of $C_{60}$-dimers are reported.

AUTHOR INFORMATION

**Corresponding Author**

*Email: keugene@bgu.ac.il

**Notes**

The authors declare no competing financial interests.

**ACKNOWLEDGMENTS**

L.A.C., A.A.A. and V.A.D acknowledge financial support by the Program F7, project 318617-FAEMCAR. All calculations were performed using the resources of the Interdepartmental Supercomputer Center (ISC) and the Lomonosov Supercomputer Complex, Moscow State University[40]. E.A.K. acknowledges financial support by the Israel Science Foundation and EC-FP7-IRSES program.